\documentclass[twocolumn,showpacs,preprintnumbers,amsmath,amssymb,prl]{revtex4}

\usepackage{graphicx}
\usepackage{dcolumn}
\usepackage{bm}
\usepackage{latexsym}
\usepackage{amstext}
\usepackage{amsxtra}

\newcommand{\bs}{\boldsymbol}

\begin{document}
\title{Quadrupole Oscillation of a Single-Vortex Condensate: Evidence for
Kelvin Modes}
\author{V. Bretin, P. Rosenbusch, F. Chevy, G.V. Shlyapnikov$^*$, and J. Dalibard}
\affiliation{Laboratoire Kastler Brossel$^{**}$, 24 rue Lhomond,
75005 Paris, France}
\date{Received October 30, 2002}

\begin{abstract}
{We study the two transverse quadrupole modes of a cigar-shaped
Bose-Einstein condensate with a single centered vortex. We show
that the counter-rotating mode is more strongly damped than in the
absence of a vortex, whereas the co-rotating mode is not affected
appreciably by the vortex. We interpret this result as a decay of
the counter-rotating quadrupole mode into two excitations of the
vortex line, the so-called Kelvin modes. This is supported by
direct observation of the wiggling vortex line. }
\end{abstract}


\maketitle

When a superfluid described by a macroscopic wave function
$\psi(\bs r)$ is set into rotation, quantized vortex lines appear,
along which the density $|\psi|^2$ is zero
\cite{Onsager,Feynman,Lifshitz}. Since $\psi$ is single valued,
its phase variation on a closed contour around a vortex line is
$2n\pi$ ($n$ is integer). Such vortices have been observed in many
systems, e.~g. superconductors \cite{Tinkham}, superfluid liquid
helium \cite{Donnelly91}, and gaseous Bose-Einstein condensates
\cite{Matthews99,Madison00,Ketterle1,Hodby01}.

A vortex line is a dynamic object. As for a classical string,
transverse vibration modes of the line can be excited. For a
classical vortex, the modes have been calculated by Kelvin in 1880
\cite{Thomson1880}. His result can be transposed to a single
quasi-linear quantum vortex with unit charge ($n=1$) in a
homogeneous superfluid. This gives the relation between the energy
$ \hbar\omega_{\rm K}$ of a quantum of the Kelvin modes (kelvon)
and its wave vector $k$ \cite{Lifshitz,Pitaevskii61,Fetter67}:
 \begin{equation}
\hbar \omega_{\rm K} \simeq \frac{\hbar^2 k^2}{2m}\;\ln(1/k\xi)
\qquad (k\xi \ll 1)\ .
 \label{dispersion}
 \end{equation}
Here $m$ is the mass of a particle of the fluid and $\xi=(8\pi
\rho a)^{-1/2}$ the healing length ($a$ is the scattering length
characterizing the binary interactions in the fluid and $\rho$ the
fluid density). Due to the Kelvin--Helmoltz theorem
\cite{theorem}, the Kelvin modes rotate always in the sense
opposite to the vortex velocity field \cite{Lifshitz,Thomson1880}.
Consequently the angular momentum of a kelvon associated with a
$n=1$ vortex is $-\hbar$ \cite{Epstein92}.

The Kelvin modes play an important role in the dynamics of
superfluids \cite{Donnelly91,Sonin87} or neutron stars
\cite{Epstein92}. In superfluid liquid helium $^4$He, they have
been observed by trapping ions in the vortex core, and exciting
them by a circularly polarized electromagnetic field
\cite{Ashton79}.

In this Letter, we present experimental evidence for Kelvin modes
excited through non-linear (Beliaev) decay of the quadrupole mode
$m=-2$. An elementary excitation of the quadrupole mode $m=-2$
decays into a pair of kelvons with wave vectors $k$ and $-k$. We
measure the corresponding increase of the decay rate with respect
to the decay of the $m=+2$ mode, for which angular momentum
conservation forbids such a mechanism. We also show pictures of
the wiggling vortex line obtained after the excitation of the
$m=-2$ mode. The spatial period of the oscillations of the line is
in good agreement with the one deduced from the generalization of
Eq.~(\ref{dispersion}) to a trapped condensate
\cite{Isoshima99,Svidzinsky00} and the energy conservation
$\omega_{-2}=2\omega_K$ ($\omega_{-2}$ is the frequency of the
$m=-2$ quadrupole mode).

We use a cigar-shaped $^{87}$Rb condensate held in an
axi-symmetric Ioffe-Pritchard magnetic trap. The atoms are
spin-polarized in the $F=m_F=2$ state. The trapping frequencies
are $\omega_z/2\pi=11.8$~Hz (longitudinal axis) and
$\omega_\bot/2\pi=98.5$~Hz (transverse plane). The temperature is
$\sim 70$~nK, corresponding to a $70$\% condensed fraction. The
number of atoms in the condensate is $1.3\times10^5$ and the
chemical potential is $40$~nK.

In order to produce in a reliable way a single vortex centered on
the $z$ axis \cite{array}, we follow the procedure outlined in
Ref.~\cite{Rosenb02}. We use an off-resonant laser beam, whose
motion is controlled by acousto-optic modulators, to superimpose a
rotating dipole potential onto the magnetic potential. This dipole
potential is asymmetric in the $x-y$ plane and reads
$m\omega_{\perp}^2 \epsilon (X^2-Y^2)/2$ with $\epsilon= 0.07$.
The axes $X,Y$ are deduced from the fixed axes $x, y$ by a
rotation of an angle $\Omega t$. The dipole potential is switched
on after the condensate formation for a period of 0.3~s. We choose
$\Omega$ close to the rotating quadrupole resonance
$\omega_\perp/\sqrt{2}$, thus creating a lattice with 5 to 7
vortices. During the next 2 seconds, the laser stirrer is blocked
and the vortex lattice decays due to a slight ($\sim 1\%$) static
anisotropy of the magnetic potential in the $x-y$ plane. The
condensate is then left with a single centered vortex. The
lifetime of this last vortex is $\sim 7$ seconds, which is much
longer than the rest of the experimental sequence. Experiments
without vortices performed for comparison follow the same
procedure, except that rotation is kept at a lower frequency to
prevent any nucleation of vortices.

The dipole potential is then used again on the single vortex
condensate to selectively excite the modes $m=+2$, $m=-2$ or their
superposition. During this excitation, the angular frequency and
deformation parameter of the dipole trap are set to $\Omega'$ and
$\epsilon'$. The duration of the excitation is denoted by $\tau$.
The state of the condensate is finally probed after a 25 ms time
of flight by simultaneous absorption imaging along the two
directions $y$ and $z$. Transverse images (imaging beam along $y$)
give indications on the behavior of the vortex line. Quantitative
information is obtained from the longitudinal images (imaging beam
along $z$). These are fitted assuming an elliptic Thomas-Fermi
profile for the spatial density of the condensate in the $x-y$
plane. We measure the size $R_{\rm l}$ and the polar angle
$\theta$ of the long axis, and the size $R_{\rm s}$ of the short
axis. $\zeta=R_{\rm l}/R_{\rm s}$ denotes the ellipticity in the
$x-y$ plane.

\begin{figure}
\centerline{\includegraphics[width=8.5cm]{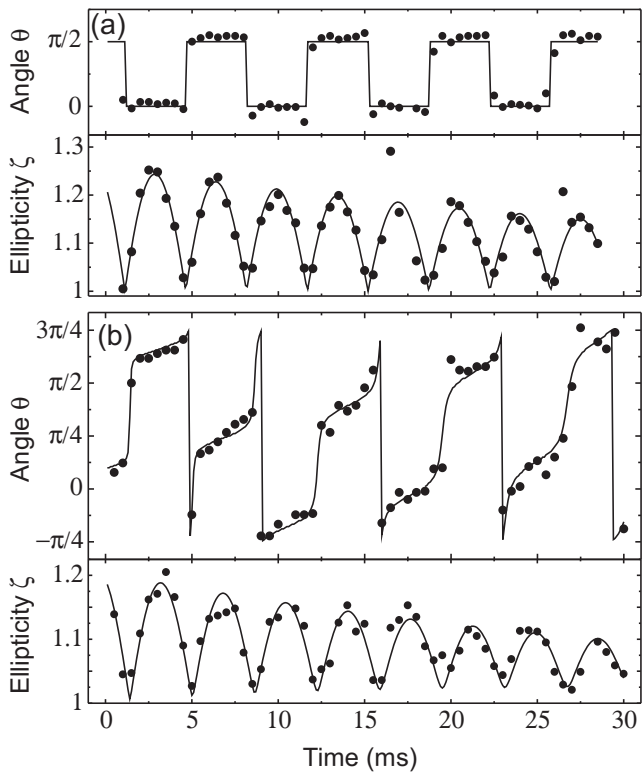}}
\caption{Angle $\theta$ and ellipticity $\zeta$ as a function of
$t$ in absence (a), and in presence of a single centered vortex
(b). In (b) the precession of the main axes observed for small $t$
increases at longer times, and the jumps in $\theta$ become
rounded indicating that the mode $m=-2$ decays faster than the
mode $m=+2$.}
  \label{figure1}
\end{figure}

The first series of experiments aims at observing the free
evolution of the two quadrupole surface modes $m=\pm 2$, in the
presence and absence of a vortex. A percussional excitation is
performed using the laser stirrer with fixed axes ($\Omega'=0$)
for a short duration $\tau=0.5$~ms $\ll \omega_\bot^{-1}$
($\epsilon'\sim 1$). This excites a superposition of $m=+2$ and
$m=-2$ modes with equal amplitudes. We then let the cloud evolve
freely in the magnetic trap for a variable time $t$ and we perform
the time of flight analysis. The quantities $\theta$ and $\zeta$
are plotted as a function of $t$ in Fig.~\ref{figure1}.

In the case where the condensate is vortex-free
(Fig.~\ref{figure1}a), the angle $\theta$ jumps periodically
between 0 and $\pi/2$, indicating that the amplitudes of the two
quadrupole modes $m=+2$ and $m=-2$ stay equal. The situation is
dramatically different in the presence of a vortex
(Fig.~\ref{figure1}b). In this case two successive regimes occur:
for short times, one still observes the quadrupole oscillation,
now with precessing axes. The precession is due to the lift of
degeneracy between the frequencies $\omega_{\pm2}$ of the two
modes $m=\pm 2$ with $\omega_{+2}>\omega_{-2}$
\cite{Dodd97,Sinha97,Svidzinsky98,Stringari}. This well known
effect is often used to measure the angular momentum of the
condensate \cite{Chevy,Cornell1,Ketterle2}. For $t>5$\,ms, the
precession rate increases and the jumps of $\pi/2$ in $\theta$
become more and more rounded. This behavior indicates that the
mode $m=-2$ decays faster than $m=+2$, which will eventually lead
to an atom cloud with non-oscillating ellipticity and constant
rotation at $\omega_{+2}$.

For a quantitative analysis, we have fitted $\theta(t)$ and
$\zeta(t)$ assuming that the amplitudes $\alpha_m$ of the modes
$m=\pm 2$ vary as
 \begin{equation}
 \dot \alpha_m + (i\omega_m +\Gamma_m)\,\alpha_m =0
 \label{evolution}
 \end{equation}
with $\alpha_{+2}(0)=\alpha_{-2}(0)$. The decay of the quadrupole
modes is modelled using phenomenological linear damping rates
$\Gamma_m$. A more refined treatment should take into account the
non-linear character of the Beliaev-type process under
investigation.

In the case of no vortex (Fig.~\ref{figure1}a), where the symmetry
between the modes $m=\pm2$ is preserved, we deduce from the fit
$\omega_{\pm 2}^{(0)}/2\pi= 142.0\pm 0.5$\,Hz, which is close to
the prediction in the Thomas-Fermi regime
$\sqrt{2}\omega_\bot\,/\,2\pi= 139.3$\,Hz \cite{StringariModes}.
For the decay rate we find $\Gamma_{\pm 2}^{(0)}=21.3\pm
1.3$\,s$^{-1}$. In the presence of a vortex (Fig.~\ref{figure1}b),
the fit gives: $\omega_{+2}/2\pi= 159.5\pm 1.0\,$Hz,
$\omega_{-2}/2\pi= 116.8\pm 1.0\,$Hz, $\Gamma_{+2}=19.1\pm
2.0\,$s$^{-1}$ and $\Gamma_{-2}=35.7\pm 4.0\,$s$^{-1}$. We note
that the two measured frequencies satisfy the sum rule
$\omega_{+2}^2+\omega_{-2}^2=4\,\omega_{\perp}^2$ \cite{Stringari}
with a good accuracy. The fit confirms the difference in the two
decay rates with the addition that the mode $m=-2$ decays faster
in the presence of a vortex ($\Gamma_{-2}>\Gamma_{\pm 2}^{(0)}$),
whereas $\Gamma_{+2} \simeq \Gamma_{\pm 2}^{(0)}$
\cite{doublefitzerovortex}.

\begin{figure}
\centerline{\includegraphics[width=8.5cm]{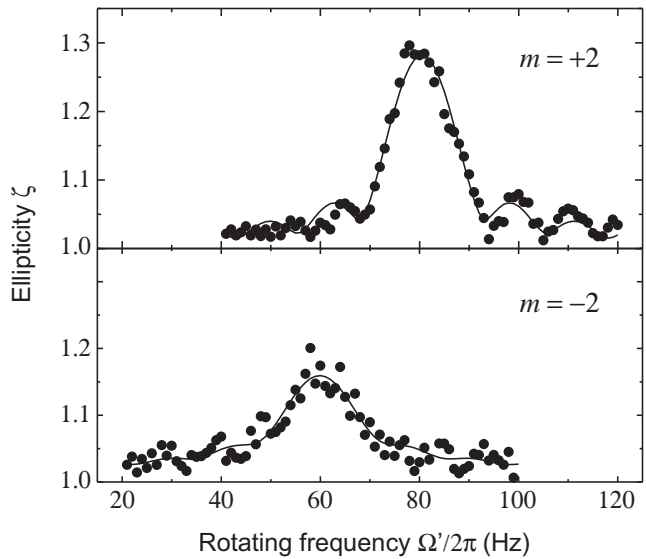}}
\caption{Resonance of the ellipticity $\zeta$ for the $m=+2$ and
$m=-2$ modes. The lines are fits deduced from the solution of
Eq.~(\ref{evolution}) with a sinusoidal drive. The resonance of
the mode $m=-2$ is broader than that of the $m=+2$, as a
consequence of a larger damping rate.}
  \label{figure2}
\end{figure}

In a second series of experiments, we perform a spectroscopy-like
study of the two modes $m=+2$ and $m=-2$ separately. For the
excitation, we set $\epsilon'= 0.025$ and apply the laser stirrer
for $\tau=40$~ms. It is rotating with $\Omega'$ either with the
same (excitation of $m=+2$) or the opposite sense ($m=-2$) with
respect to the vortex. We perform the usual time of flight and
imaging, and measure the ellipticity $\zeta$ as a function of
$\Omega'$. We observe a clear resonance for each mode
(Fig.~\ref{figure2}). The two resonances occur at different
central frequencies according to the already mentioned lift of
degeneracy. The key feature of the present study is that the width
of the $m=-2$ resonance is significantly larger than that of
$m=+2$. Hence, as seen before, the  vortex causes larger damping
for the counter-rotating surface wave than for the co-rotating.

To analyze our data, we assume that the amplitude of the relevant
mode varies according to Eq.~(\ref{evolution}) with a driving term
$A\, e^{-2i\Omega't}$ on the right hand side. This assumption
leads to a good fit to the data (Fig.~\ref{figure2}) and yields
$\omega_{+2}/2\pi= 161.0\pm 1.0\,$Hz, $\omega_{-2}/2\pi= 119.8\pm
1.0\,$Hz. These frequencies are in excellent agreement with those
obtained by the percussional excitation. The $\sim 2$~Hz shift is
due to the increase of the trapping frequency by the laser
stirrer. The fitted damping rates are $\Gamma_{+2}=24\pm
5\,$s$^{-1}$ and $\Gamma_{-2}=57\pm 10 \,$s$^{-1}$. These are
slightly larger than those obtained from the percussional
experiment, possibly due to an additional heating caused by the
stirrer. However, the main feature $\Gamma_{-2} \sim 2\;
\Gamma_{+2}$ remains valid.

We now discuss the possible physical origin of this increased
damping of the $m=-2$ mode. The starting point of our analysis is
the behavior of the uncondensed part of the gas. Is it still
rotating when we perform the percussional or spectroscopic study?
If it rotates, the symmetry between the two senses of rotation is
broken and the co-rotating mode experiences less friction than the
counter-rotating mode. In \cite{Griffin}, it has been shown that
in this case $\Gamma_{-2} > \Gamma_{\pm 2}^{(0)}$ and $\Gamma_{+2}
< \Gamma_{\pm 2}^{(0)}$. Moreover one gets $\Gamma_{-2} +
\Gamma_{+2}\simeq 2\Gamma_{\pm 2}^{(0)}$. This latter relation
does not correspond to our observations. Although we do observe an
increase of $\Gamma_{-2}$ when a vortex has been nucleated, we do
not observe a corresponding reduction of $\Gamma_{+2}$. From this
we infer that the uncondensed fraction does not rotate
significantly in our experiment. Furthermore, no such rotation is
expected for our experimental conditions. The $\sim 1\%$ static
trap anisotropy, which barely affects the behavior of the
condensate, rapidly damps the rotation of a non-condensed gas
\cite{Guery, Cornell2}. The thermal cloud thus experiences two
competitive forces in the bare magnetic trap: a rotational drive
from the velocity flow of the condensate, and a friction from the
static trap anisotropy. Unlike in the case of a complete vortex
array \cite{Fedichev}, the rotational drive due to a condensate
with a single vortex is small compared to the friction term.
Therefore the rotation of the thermal cloud is expected to be
negligible.

Assuming now that the uncondensed gas is at rest, the other
possible reason for a faster damping of the $m=-2$ mode is the
existence of a decay channel open for $m=-2$ and closed for
$m=+2$. The natural candidate for this mechanism is the Beliaev
conversion of the $m=-2$ mode into two Kelvin modes $k$ and $-k$,
followed by the kelvon decay into another kelvon with a slightly
smaller momentum and a phonon. Since the Kelvin excitations have
negative angular momentum with respect to the vortex charge, this
mechanism is specific to the $m=-2$ mode. A recent theoretical
analysis of this decay process, taking into account both Beliaev
and Landau damping, has led to damping rates in good agreement
with those that we  measure experimentally \cite{Machida}.

\begin{figure}
\centerline{\includegraphics[width=8.5cm]{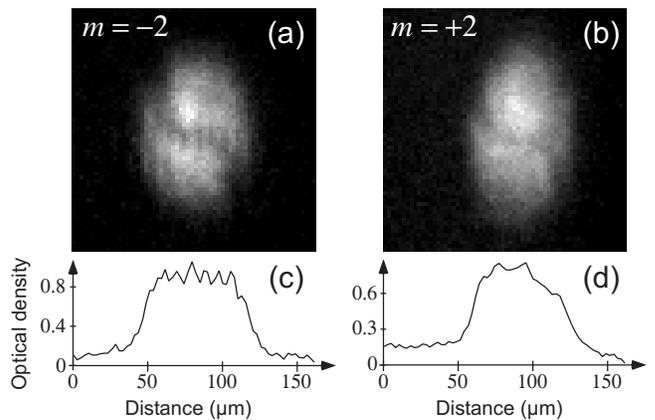}}
\caption{Transverse images of the condensate after a 33 ms
resonant excitation of the $m=-2$ mode (a) and the $m=+2$ mode
(b). The  picture on the left shows a periodic structure
superimposed on the vortex line. The horizontal density profiles
(c,d) have been obtained by averaging over three vertical pixels
around the center of the vortex line.}
  \label{figure3}
\end{figure}

To support the interpretation of decay into Kelvin modes, we
present time-of-flight images of the vortex line. The images are
taken immediately after the excitation of the $m=-2$ or $m=+2$
mode by the spectroscopy-like method. For each mode we choose
$\Omega'$ at the center of the resonance and apply the probe for
$\tau= 33$\,ms. Fig.~\ref{figure3} shows two condensates in the
transverse view. The left picture (a) was taken counter-rotating
($m=-2$), the right picture (b) co-rotating ($m=+2$). Below are
shown the horizontal density profiles taken at the center of the
vortex line (c,d).

The vortex line is visible in both images with good contrast. It
has the form of an unfolded ``N" \cite{Rosenb02}. After excitation
of the $m=-2$ mode, the atom density presents a periodic structure
apparent as vertical stripes in the vicinity of the vortex line
(Fig.~\ref{figure3}a,c).  A Fourier analysis of the density
profile reveals a peak centered at $k_0=0.70$~$\mu$m$^{-1}$ with a
full width at half maximum $\Delta k_0\simeq 0.15$~$\mu$m$^{-1}$.
The area of the peak, which can be observed reproducibly for
similar experimental conditions, is 6 standard deviations above
noise. On the contrary, such a periodic structure never appears in
the images corresponding to $m=+2$ (see e.g.
Fig.~\ref{figure3}b,d).

This observation supports the proposed decay mechanism of the
$m=-2$ mode into a pair of Kelvin modes. At any given time, the
excited vortex line is expected to have an almost sinusoidal shape
with a spatial period of $2\pi/k$ along the $z$ axis. It is
located in a plane containing the $z$ axis and rotating around
this axis in the sense opposite to the vortex. At the center of
the condensate, the healing length is $\xi\sim 0.3\;\mu$m. Using
Eq.~(\ref{dispersion}) and energy conservation, we find
$k=0.8$~$\mu$m$^{-1}$. Assuming that during the time of flight,
the expansion factor $\Lambda_z$ along the $z$ axis is the same
for the shape of the vortex line as it is for the longitudinal
size of the condensate ($\Lambda_z\sim 1.3$ \cite{Castin96}), we
recover the measured wave vector $k_0$ within $\Delta k_0$. This
gives $\sim 9$ oscillation periods over the $L\sim 70\;\mu$m
length of the condensate, which is consistent with the observation
in Fig.~\ref{figure3}a. With the quantization condition $k={\cal
N}\pi/L$ for the vortex line, we get ${\cal N}\sim 18$. The
splitting between the adjacent Kelvin modes is then $\sim
20\,$s$^{-1}$, so that we have one or two Kelvin modes within the
width $\Gamma_{\pm 2}^{(0)}$ of the quadrupole resonance. We note
that a detailed treatment of the relation between the observed
density modulation and the oscillations of the vortex line remains
to be done. The principle of such a treatment may follow that of
Ref.~\cite{Ertmer}, where phase fluctuations due to a short
coherence length in a quasicondensate appear as density
fluctuations after a time of flight.

In conclusion, we have observed a strong difference between the
damping rates of the co-rotating ($m=+2$) and counter-rotating
($m=-2$) quadrupole modes of a single vortex condensate. We
explain this difference as due to the decay of the $m=-2$ mode
into Kelvin excitations of the vortex line. This is confirmed by
images of the line, which show wiggling only for counter-rotating
excitation.

{\acknowledgments We thank K. Madison for participation in earlier
stages of this experiment, and L. Carr and Y. Castin for useful
discussions. We also thank K. Machida for sending us his results
prior to publication. P. R. acknowledges support by the
Alexander-von-Humboldt Stiftung and the EU (contract number HPMF
CT 2000 00830). This work is partially supported by CNRS,
Coll\`{e}ge de France, R\'{e}gion Ile de France, DGA, DRED and EU
(TMR network ERB FMRX-CT96-0002).}

\end{document}